\documentclass[usenatbib]{mn2e}
\usepackage{epsfig}
\usepackage{amsmath}
\title[Scatter in the X-ray scaling laws]{An analytic investigation of the scatter in the
integrated X--ray properties of galaxy groups and clusters.}
\author[Balogh \etal]{Michael L. Balogh$^{1}$, Arif Babul$^{2}$, G. Mark
  Voit$^{3}$, Ian G. McCarthy$^{2,4}$, \newauthor
Laurence R. Jones$^{5}$, Geraint F. Lewis$^{6}$, Harald Ebeling$^{7}$\\
$^{1}$Department of Physics, University of Waterloo, Waterloo, ON,
Canada N2L 3G1, email: mbalogh\@@uwaterloo.ca\\
$^{2}$Department of Physics and Astronomy, University of Victoria,
Victoria, BC, Canada V8P 1A1\\
$^{3}$Department of Physics and Astronomy, Michigan State University,
East Lansing MI 48824 USA\\
$^{4}$Department of Physics and Astronomy, University of Durham, Durham, UK, DH1 3LE\\
$^{5}$Department of Physics and Astronomy, University of Birmingham,
Birmingham, UK B15 2TT\\
$^{6}$Institute of Astronomy, School of Physics, A29, University of
Sydney, NSW 2006, Australia\\
$^{7}$Institute for Astronomy, 2680 Woodlawn Drive, Honolulu, Hawaii 96822 USA\\
}
\date{\today}
\def\etal{{ et al.\thinspace}}
\def\gtrsim{\mathrel{\raise0.35ex\hbox{$\scriptstyle >$}\kern-0.6em
\lower0.40ex\hbox{{$\scriptstyle \sim$}}}}
\def\lesssim{\mathrel{\raise0.35ex\hbox{$\scriptstyle <$}\kern-0.6em
\lower0.40ex\hbox{{$\scriptstyle \sim$}}}}

\begin{document} 
\maketitle
\begin{abstract}
We revisit the scaling relationships between the dark matter mass and
observed X--ray luminosity and temperature of galaxy clusters and
groups in the local Universe.  Specifically, we compare
recent observations with analytic models of the intracluster medium in which
the gas entropy distribution has been shifted by a variable amount,
$K_\circ$, to investigate the origin of the scatter in these scaling
relations, and its influence on the luminosity and temperature
functions.  We find that variations in halo
concentration or formation epoch (which might determine the time
available for low entropy gas to cool out) are insufficient to explain
the amount of scatter in the mass--luminosity relation.  Instead, a
range of entropy 
floors at a fixed halo mass, spanning approximately $\sim
50$ keV~cm$^2$ to $\sim 700$ keV~cm$^2$, is required to match the data.
This range is likely related to the variance in heating and/or cooling
efficiency from halo to halo.  We demonstrate that these models
are consistent with the observed temperature and luminosity functions of
clusters, with a normalization of $\sigma_8\sim 0.8$ in agreement with
WMAP measurements (for
$h=0.7$ and $\Omega_m=0.3$); in particular the scatter in
the mass--luminosity relation has an important influence on the shape
of the luminosity function, and must be accounted for to provide a
consistent result.
Finally, we present predictions for the redshift evolution of these scaling
relations and luminosity/temperature functions.  
Comparison with recent data at $z<0.7$ shows reasonable agreement with
a model that assumes a median entropy floor of $K_\circ=200$ keV~cm$^2$.
When observations are extended to group
scales ($kT\lesssim 1$keV), this evolution will have the potential to discriminate between an
entropy floor that is independent of redshift (for example, in a
preheating scenario) and one that depends on the cooling time of the halo.
\end{abstract}
\section{Introduction}\label{sec-intro}
The X-ray properties of clusters are tracers of both the gravitational
potential and the thermodynamic history of the gas.  Since the mass is
likely dominated by collisionless dark matter, for which we have a
well--developed theory \citep[e.g.][]{EvrardHV}, we might hope to learn
about the relevant baryonic physics through detailed X-ray observations.
In particular, clusters obey fairly well defined scaling relations
between mass and X-ray temperature (M-T) and mass and X-ray luminosity (M-L).
Although the slope and normalization of these
relations relative to model predictions have been studied extensively
\citep{ES91,Markevitch,HMS,NMF,APP}, little attention has been paid to their
intrinsic scatter \citep[but see][]{RTK,KTJP,GPS4}.

The presence of scatter in the M-L and M-T relations suggests intrinsic
variations in the structure of clusters, which may be due either to variations in the 
   dark matter distribution of the halos themselves, and/or in the gas
   properties.  In particular, the scatter is likely to be driven by
   variations in the core properties of clusters, since the scatter is known to be significantly reduced
   if the central regions are excluded from the analysis
   \citep[e.g.][]{Markevitch}.

In the case of the dark matter component of relaxed clusters, there are
   indications that either (or both) the halo concentration parameter
   and the inner slope of the halo profile varies from cluster to cluster,
   perhaps due to cosmic environmental effects such as the extent of tidal
   torquing the dark matter experiences during collapse or the merger history of
   the halo \citep[e.g.][]{Jing00,Bullock01,Wechsler02,ZMJB,WBD04}.
   The amount of substructure and dynamical state of the dark matter
   will also vary and depend upon the merging history of the halo.

There are also reasons to expect substantial variation in the gas
properties of clusters, independently of their dark matter distribution.
Models that seek to account for the mean
   M-L and M-T relations require some form of entropy modification
   in the central regions of the systems \citep{TMKS,V+03}.   Pure heating models have been
   very successful in explaining the average trends of these scaling
   relations, especially when the heating targets the lowest entropy
   gas \citep[hereafter BBLP]{K91,entropy,Babul2}.  It seems likely that the efficiency of whatever
   physical mechanism is responsible for the heating (e.g. heat transport, AGN energy
   injection, etc) will vary from cluster to cluster.  
In this case, one expects both cooling and the feedback it triggers to
eliminate gas below an entropy threshold that depends on halo mass and redshift \citep[][hereafter
   VBBB]{VB01,Voit}, although some gas may exist below this threshold
   if it is in the process of cooling out.
   In a realistic cooling model \citep[e.g.][]{McCarthy-cooling},
   scatter can be
   introduced by appealing to a range in time available for cooling in
   each halo.  Despite this, McCarthy et al. show that such a range
   alone cannot account for the scatter in the M-L relation; it is also necessary to
   introduce variations in an initial heating level.  However, they do
   not consider the effect of variations associated with the underlying
   dark matter potential, which will contribute some scatter
   independently of the gas entropy distribution.

Shorter--lived changes to the equilibrium
   temperature and luminosity of a cluster may also be associated with
   merger events \citep[e.g.][]{RT02,RSR,RTK}.  Departures
   from equilibrium in the potential can change the luminosity, but have little effect on the gas temperature
   \citep{RTK}. However, the shocks associated
   with mergers can have a temporary but significant influence on
   both the temperature and luminosity of the gas \citep{RT02,RTK}.  

Recently, semi-- and fully--numerical simulations which include both cooling and
feedback from star formation have been shown to produce clusters with
X--ray properties that scale with mass in a way that is in 
reasonable agreement with the observations
\citep[e.g.][]{M+01,TMKS,Borgani02,V+03,KTJP,RTK,Borgani+05,OBB}.  
In this paper we will re-examine the observed M-L and M-T relations, focusing
on the scatter in these relations and how it relates to the expected
variation in the underlying physical processes.  
The data we use are uncorrected for any cooling-core
component, since it is precisely this core region that interests us. 
We will compare these
observations with analytic, hydrostatic models which allow
modifications to the entropy distribution of the gas (BBLP,VBBB), to determine the range of
model parameters that are required to reproduce the observed scatter.
Although the hydrostatic nature of these models means they are not
ideally suited to explore in detail the effects of active cooling or heating in
clusters, the range of model parameters required to match the data can
be related indirectly to these processes.
We will also make a
self-consistent comparison with the temperature and luminosity
functions, which provide an independent test of the models assuming the
dark matter mass function is known \citep{EvrardHV}.  Finally, we will
present the redshift evolution 
   of all these observable quantities to put further constraints on the
   model parameters.   

 Throughout this paper we use a
cosmology with $\Omega_m=0.3$, $\Omega_\Lambda=0.7$ and $H_\circ=70$~km~s$^{-1}$~Mpc$^{-1}$.
The theoretical models are described in \S~\ref{sec-models}, and the
comparison with observed X-ray properties is presented in
\S~\ref{sec-MLT}.  Predictions for the evolution of these models, and
some comparison with early data, are given in \S~\ref{sec-evol}.  We
summarize our conclusions and discuss the implications and limitations of our findings, 
in
\S~\ref{sec-discuss}.  

\section{The models}\label{sec-models}
\subsection{Dark matter profile shapes}\label{sec-darkmatter}
The average theoretical shapes of dark matter halos are well motivated by
N-body simulations to have a form given by 
\begin{equation}\label{eqn-nfw}
\rho\propto r^{-n_1}(1+c_{200}r)^{-n_2}
\end{equation}
\citep[][NFW]{NFW3}, where $n_1$, $n_2$, and $c_{200}$ are fitting
parameters. The primary determinant of a halo's
structure is its virial mass, $M_{\rm vir}$.   This is commonly defined,
using
spherical collapse models, as the mass within a fixed
overdensity $\Delta$ that depends on cosmology and redshift; for  $\Lambda CDM$, 
$\Delta\sim 100$ at $z=0$ \citep{ECF}.  However, observations are more typically made at $\Delta=200, 500$ or
larger.  The radius and mass corresponding to the overdensity $\Delta$ will be
denoted $R_\Delta$ and $M_\Delta$, respectively.  

High resolution simulations show that relaxed clusters have remarkably
uniform profiles.  We will assume an NFW profile with $n_1=1$
and $n_2=2$ as our fiducial model.  The range of values of these
parameters reported in the literature
\citep[e.g.][]{NFW-Moore,Lewis-sim} appear to be mostly due to differences
in resolution and the fact that the fitting formula is not a perfect
description of the profile \citep{HayashiII}.  Within a given
simulation, an indication of the amount of variation in halo shapes is
best given by the distribution of concentration parameters, $c_{200}$.  This parameter has a systematic
dependence on mass and redshift
\citep[e.g.][]{ENF,Bullock01,Powerthesis,Wechsler02}; we will take the
parameterization of \citet{ENS}, assuming $\sigma_8=0.8$.  Most
importantly for our purposes, $c_{200}$ shows considerable scatter at fixed mass and
redshift; \citet{Dolag04} have
recently shown that the distribution of
$c_{200}$ values in simulated clusters (selected only based on an
overdensity criterion) is approximately log-normal with a width
that is nearly independent of mass.  This scatter is at least partly
due to the presence of substructure, triaxiality, and departure from
equilibrium in the sample of simulated clusters chosen from the simulations.
We will therefore consider a range of concentrations corresponding to
the $\pm3\sigma$ range predicted from this distribution.  Although
different values of $\sigma_8$ will change the value of $c_{200}$ at
fixed mass by a small amount, it will not have an important effect our
discussion in this paper, which is based on the variation in $c_{200}$
and not its absolute value. 

\begin{figure}
\leavevmode \epsfysize=8cm \epsfbox{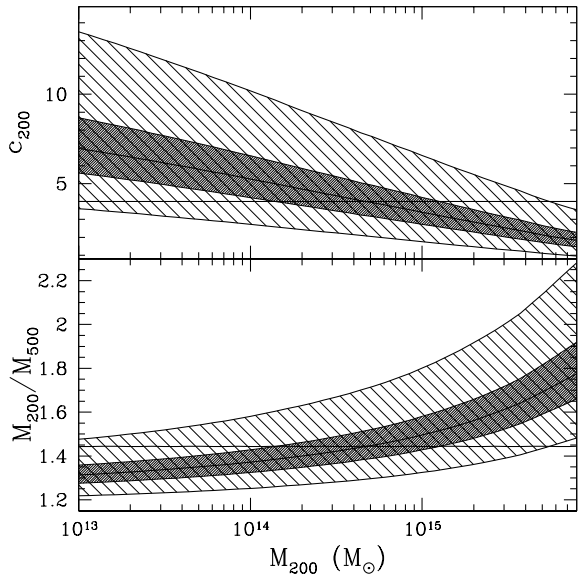} 
\caption{
{\bf Top panel: } The relation between $c_{200}$ (concentration) and $M_{200}$ as a function of $M_{200}$
is shown as the {\it shaded region}, where the distribution is due to
the $1-\sigma$ (heavy shading) and $3-\sigma$ (light shading) distribution of concentrations from \citet{Dolag04}.  The
horizontal line shows a fixed concentration of 4, for reference.
{\bf Bottom panel: }
The relation between $M_{500}$ and $M_{200}$ as a function of $M_{200}$
is shown as the {\it shaded region}, assuming the same distribution of
concentrations above.  The horizontal, solid line shows the result for a model with a fixed
concentration, $c_{200}=4$.
\label{fig-Mvir}}
\end{figure}

In Figure~\ref{fig-Mvir} we show how the concentration and its scatter
depend on dark halo mass $M_{200}$ in our model.  The average
concentration and its scatter both decline with increasing mass.  In
the bottom panel we show how this affects the ratio
 $M_{200}/M_{500}$.
The scatter in concentration at a fixed mass
corresponds to a $\sim 10$ per cent scatter in $M_{200}/M_{500}$.  The
dependence of concentration on mass means the slope of measured
correlations between observables (like X-ray temperature and
luminosity) and mass will depend on which mass is used.  Throughout
this paper we will present our results as a function of $M_{200}$, and
this relation can be used to deduce the corresponding value of $M_{500}$.

\subsection{Entropy distributions}\label{sec-VBBB}
The shape of the gas profile in a cluster of given mass is determined by the entropy
distribution of that gas, which is sensitive to its thermodynamic
history.   We adopt the common \citep[e.g.][]{Ponman} redefinition of
  entropy as $K \equiv kT_e n_e^{-2/3}$, where $T_e$ and $n_e$ are the
  electron temperature and density, respectively.  This is related to
  the thermodynamic entropy by a logarithm and an additive constant, and
  is given in units of keV~cm$^{2}$.  We will usually quote this
  quantity in dimensionless units, relative to $K_{100}=100$keV~cm$^2$.

We use the formalism of VBBB to compute the hydrostatic equilibrium gas
distributions under different assumptions about the dark matter
potential and the thermodynamic history of the gas.  Once the halo potential is specified, the
entropy distribution and appropriate boundary conditions are all that
are required to fully describe the gas density and temperature
profiles.  We start with an initial profile where the gas density
traces the dark matter density, and solve for the temperature profile
needed to satisfy the assumption of hydrostatic equilibrium.  This is a
good approximation at large radii, where the entropy scales
approximately as $K\propto r^{1.1}$, as found from analytic modelling
and numerical simulations
\citep[][]{Lewis-sim,TN01,Voit03,MFB}. The gas distribution at smaller
radii is dominated by the entropy modifications that we discuss below,
so the choice of initial profile is relatively unimportant.
The normalization of this initial, unmodified model, is chosen by assuming the gas fraction within $R_{200}$
is equal to the global baryon fraction of 12.9 per cent \citep[for $h=0.7$
and $\Omega=0.3$,][]{BBN2,WMAP_short}.  

We will explore modifications to this default profile, in the form of the
shifted--entropy models of VBBB.  In this case, the entropy
distribution is shifted by an additive constant, which provides a good
approximation to pre-heated models.  This entropy shifting causes the gas to expand beyond
$R_{200}$; following
VBBB we therefore choose as our pressure boundary condition the accretion
pressure at the maximum radial extent of the gas.  The amount by which the entropy
distribution is shifted will be left as a free parameter; for our base
model we use $K_\circ=2K_{100}$, which is known to provide a reasonable
match to the median global scaling relations of clusters \citep{McCarthy-cooling}.

\begin{figure}
\leavevmode \epsfysize=8cm \epsfbox{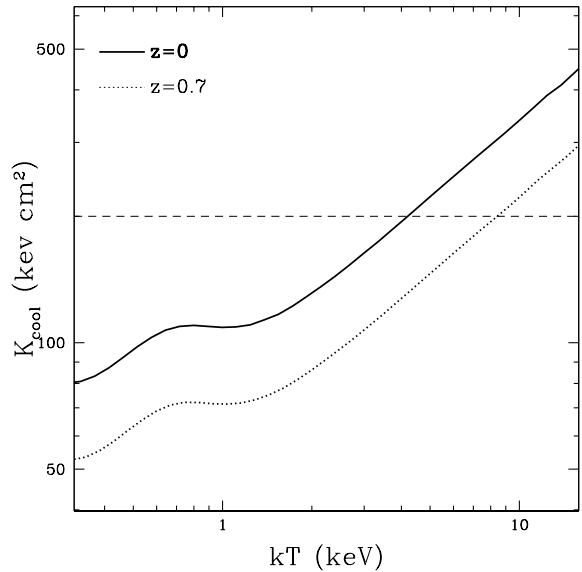}
\caption{The entropy $K_{\rm cool}$ corresponding to the maximum
  entropy for gas that
  can cool in a Hubble time, as a function of halo virial temperature
  and redshift. 
The horizontal,
  {\it dashed line} shows the $K_{cool}=2K_{100}$ line for reference.
\label{fig-Kcevol}}
\end{figure}
We will also consider the case where $K_\circ$ is set by the entropy of
gas that can efficiently cool in time $t$.  This entropy, which we will
call $K_{\rm cool}$, depends on mass and redshift \citep{VB01}, but
also on the time available for cooling \citep{McCarthy-cooling}.  Figure~\ref{fig-Kcevol}
shows the maximum value of $K_\circ$ predicted in this cooling--based
model, calculated assuming the gas can cool for a Hubble time.  
The mean value of $K_{\rm cool}$
decreases with increasing redshift due
primarily to the decrease in time available for cooling.  Of course,
simply shifting the entropy distribution by a constant value 
is not an accurate representation of the effects of radiative cooling.
In reality, an inner entropy gradient (rather than a floor) may be
expected, with some gas below the entropy threshold
\citep[VBBB;][]{McCarthy-cooling}.  This can be particularly important
for clusters, where the central entropy in a more realistic model can
drop well below the value of $K_{\rm cool}$. Furthermore, 
the time available for cooling will likely vary from cluster
to cluster, but must always be less than a Hubble time.  Both of these
effects tend to reduce the value of $K_\circ$; thus our prediction in
Figure~\ref{fig-Kcevol} represents a strict maximum value for this quantity.

It is important to note that the entropy modifications that we 
explore in the present study are all confined to the cluster core.  Therefore, 
our approach is to effectively bracket the range of central entropy levels 
required for the heating-based and cooling-based models to explain the scatter 
in the observational data.  Both models implicitly assume that the entropy 
distribution at large radii is essentially identical to that found in 
clusters formed in cosmological numerical simulations.  Gas at large
distance from the centre is not easily affected by cooling or
non-gravitational heating processes once the cluster is assembled \citep[e.g.][]{Borgani+05,OBB}.
However, heating of the gas before it is accreted into a cluster can
smooth out the density distribution of infalling gas, and this
increases the entropy jump at the accretion shock \citep{Voit03,Borgani+05}.  In this case, the
entropy distribution at large radii will be larger than that we have
assumed, and a lower value of central entropy will be required to
explain the observations. This underscores the need for detailed comparisons 
of theoretical models to the spatially-resolved entropy distributions of 
large, representative samples of clusters.  At present, however, only a 
relatively small number of clusters have accurately-determined entropy 
profiles from {\it Chandra} and {\it XMM-Newton} data.  Analysis of this small 
dataset seems to confirm that for high mass clusters ($kT \gtrsim 4$ keV) 
the entropy distribution at large radii does indeed trace the spatial
distribution  and normalization
predicted by `adiabatic' hydrodynamic simulations
\citep{McCarthy-cooling}.
This conclusion is strengthened by the fact that for high mass clusters there 
also appears to be excellent agreement between the observed projected 
temperature profiles at large radii and those predicted by hydrodynamic 
simulations \citep[e.g.,][]{DM02, L+02, V+04}.  Thus, our estimates of 
$K_\circ$ for such systems should be robust.  However, for cooler systems, 
there are preliminary 
indications of excess entropy at large radii \citep[e.g.,][]{PSF, PA04}.  As 
such, our estimates of $K_\circ$ for low temperature systems should be treated 
with some caution.  In this paper, our conclusions rest primarily on
the data for the high mass clusters.

\subsection{Prediction of observable quantities}
To compute
X-ray observables from the analytic gas profiles we use the cooling functions of
\citet{RCS} for gas with one third solar metallicity.  To avoid the
need to make bolometric corrections to the data, which depend on an
accurate measurement of the gas temperature, we compute the model
luminosities within the observed {\it ROSAT} energy bands 0.1--2.4 keV
and 0.5--2.0 keV.  Total 
luminosities are obtained by integrating out to a minimum surface
brightness of $1\times 10^{-15}$
ergs~s$^{-1}$~cm$^{-2}$~arcmin$^{-2}$, similar to that of the WARPS
survey  \citep{WARPSI}, unless stated otherwise\footnote{We note that \citet{WARPSI} attempt to correct for
flux below the surface brightness limit. However, the flux correction to the
data (typically a factor $\sim 1.4$) is an underestimate at the lowest
luminosities,  since it assumes a surface brightness profile slope 
of $\beta$=0.67, appropriate for high and moderate luminosity 
clusters but not for low luminosity groups. Thus,
the lowest luminosity data points may still systematically underestimate 
the luminosities by a factor $\lesssim$2.}. 
The sensitivity of
our results to this limit are explicitly noted. 

The model temperatures we compute are emission-weighted by the
0.1-2.4 keV luminosity, again excluding regions below the WARPS surface
brightness limit; however the choice of energy band and surface
brightness limit have a negligible effect on the calculated
temperatures for our purposes.  Simulations suggest that spectral
temperatures, as measured observationally, can be 10--20 per cent
higher than emission--weighted temperatures \citep{ME01,Rasia+04,Vik05}.  For
relaxed clusters, the difference is probably closer to the lower end
($\sim 10$ per cent) of this range \citep{Rasia+04}.  However, this 
systematic error is not of major concern for the present study, as the 
statistical errors associated with the observed temperature, mass, 
and luminosity of a given cluster are typically twice as large as this.  
For example,
in \citet{McCarthy-cooling} we derived similar constraints on 
the parameters of heating and cooling models of the intracluster medium
(ICM) from
independent analyses of the M-L and 
luminosity--temperature (L-T) relations, which suggests that small systematic errors in
temperature measurements do not have a noticeable influence on our results.

Our primary source of data is the HIFLUGCS
cluster sample \citep{HIFLUGCS}.  This survey is an X-ray
flux-limited sample of nearby clusters based on the
{\it ROSAT} All Sky Survey, in which cluster masses are determined from the
density profiles, assuming hydrostatic equilibrium and isothermal
temperature profiles.
It is common practise when considering X-ray scaling relations to use
temperatures and luminosities that are corrected for a cooling flow
component, and it is known that this reduces the scatter in these
relations \citep[e.g.][]{Markevitch}.  
However, since it is precisely this scatter that is the focus of our study,
we will present all the data as observed, without this correction.  We
take these raw cluster luminosities, in the 0.1--2.4 keV band, 
from the \citet{HIFLUGCS} catalogue.
Since many of the temperatures and masses in this catalogue have been
subjected to a significant cooling-flow correction, we will
only keep the $\sim 80$ clusters (out of 106) for which uncorrected temperatures
are available, from the catalogue of \citet{Hornerthesis}.  The
dynamical mass estimates, $M_{200}$ and $M_{500}$, are derived from the
gas temperature, assuming hydrostatic equilibrium.  To be fully
consistent, we make a small adjustment to these masses ($M\propto T$) so they agree
with the original (uncorrected) temperatures.  

\section{Scatter in the M-T and M-L scaling relations}\label{sec-MLT}
\subsection{The M-L relation}
\begin{figure}
\leavevmode \epsfysize=8cm \epsfbox{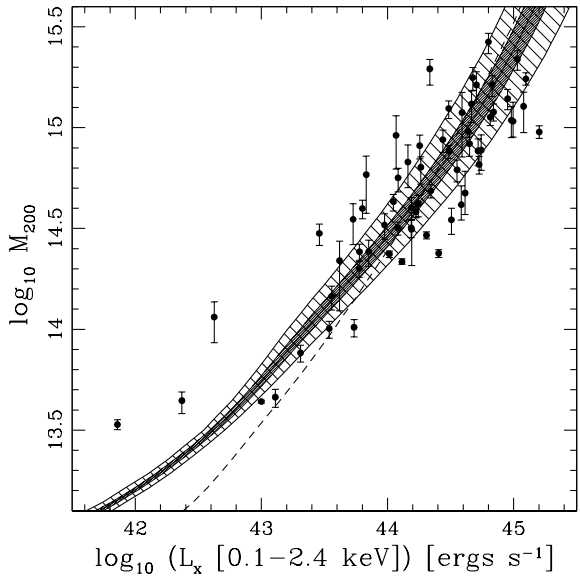} 
\caption{The relation between X-ray luminosity
  in the 0.1-2.4 keV band and halo mass $M_{200}$.
The {\it filled circles}  are local ($z<0.2$) data from the HIFLUGCS sample
  \citep{HIFLUGCS}. 1$-\sigma$ error bars on the masses are shown; only
  clusters with relative errors of $<50$ per cent are plotted.  The
  {\it shaded region} shows the shifted entropy model with
  $K_\circ=2K_{100}$ and a 1$\sigma$ (heavy shaded region) and
  3$\sigma$ (lighter shaded) range of halo concentrations.  The
  {\it dashed line} represents the model with $K_\circ=K_{\rm cool}$,
  using the most probable value of $c_{200}$ at each mass.
\label{fig-ML}}
\end{figure}

In Figure~\ref{fig-ML} we present the M--L relation from the data of \citet{HIFLUGCS}, excluding
those clusters with mass uncertainties greater than 50 per cent.   Note
that the mass we plot here is $M_{200}$.  Observationally, $M_{500}$
can be more precisely determined, and this can reduce the
purely observational scatter in this Figure \citep{RTK}.  However, when
we construct the temperature and luminosity function in \S~\ref{sec-Tfunc}
we will have to use $M_{200}$, since this was the mass used to derive
the mass function from numerical simulations \citep{EvrardHV}.  For consistency,
therefore, we have used $M_{200}$ throughout the paper.
We show both observed and model luminosities in the
0.1--2.4 keV band, to minimize errors in bolometric
corrections to the observations.  The limiting surface brightness cut
applied to the models (see \S~\ref{sec-models}) results in a
significant reduction of luminosity only below $M_{200}\sim 10^{13.5}
M_\odot$.  The shifted--entropy model reproduces the slope and
normalization of the relation well.  However,
the scatter in the data is much larger than expected from the
observational uncertainties, as has been noted before \citep[e.g.][]{FCEM,Markevitch}.

Our calculations show that the observed scatter is much larger than can be
expected from variations in halo concentration.  This is shown by the shaded region in
Figure~\ref{fig-ML}, which represents the $1-$ and $3-\sigma$ range of luminosities predicted at a
given mass, from the dispersion in simulated cluster concentrations alone.
Approximately 25 per cent of the observed clusters lie well outside
the 3$\sigma$ range resulting from variations in halo concentration.
Even though our models assume spherical, smooth, virialized halos,
these effects are partly accounted for
by the variation in concentration parameter, which is determined from
simulated clusters that are clumpy, non--spherical and in a variety
dynamical states.  Furthermore, the predicted scatter in the scaling relations is
not significantly larger in models which consider more realistic
potential shapes \citep{RTK,OBB}.  
It is therefore unlikely that the observed scatter in this relation
can be entirely attributed to variations in the shape of the dark matter distributions.

Some of the scatter in the M--L relation could be due to short--timescale
events like mergers, which cause changes in both the luminosity and
temperature of the gas \citep[e.g.][]{RT02,RSR,RTK}.  However, it has
been shown that these
changes tend to move galaxies along the L--T relation, and do not
contribute significantly to its scatter; since the observational
scatter in the L--T relation is comparable in
magnitude to that in the M--L relation of Figure~\ref{fig-ML}
\citep{FCEM,Markevitch}, there must be an important source of scatter
other than mergers.
Furthermore, as major mergers in massive clusters are expected to have
been relatively rare in the past $\sim 2$ Gyr \citep[e.g.][]{KW,LC94}, and the luminosity and temperature boosts
typically last for $\sim 0.5$ Gyr or less following a major merger,
these events are relatively rare and
unlikely to be responsible for all the observed scatter in an unbiased
cluster sample.  

\begin{figure}
\leavevmode \epsfysize=8cm \epsfbox{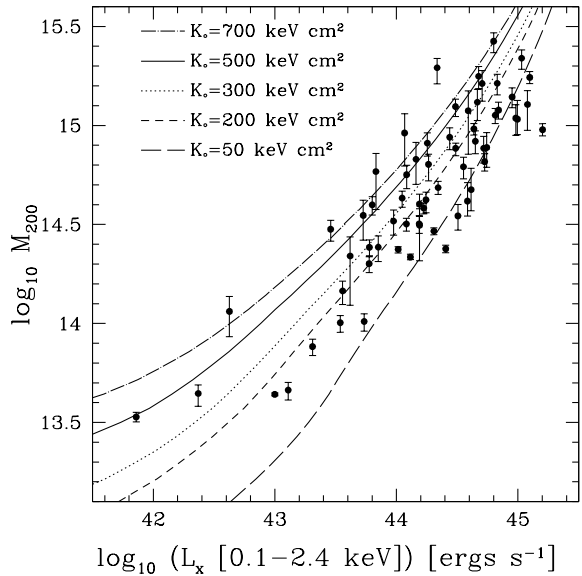} 
\caption{The mass--luminosity relation, with data as in
  Figure~\ref{fig-ML}, and five models of fixed concentration but
  different entropy thresholds, as labelled.  
\label{fig-ML2}}
\end{figure}
On the other hand,
the X--ray luminosity of a cluster of given mass is very sensitive to the entropy floor level, as shown 
in Figure~\ref{fig-ML2}.  A range of entropy levels 0.5--5 $K_{100}$ approximately
covers the scatter in the observations, at high masses.   For a few
clusters even higher levels of $K_\circ\approx 7K_{100}$ are
required to match their low luminosities, something that was also
observed in our earlier comparison with Sunyaev--Zeldovich measurements
\citep{McCarthy-SZ2}.  This is a remarkably large range in central entropy
levels, and thus substantial variations in heating or cooling
efficiency must exist from cluster to cluster, even if some of the observed scatter can be attributed to
substructure, departures from equilibrium, or merger--induced shocks.

In the cooling--based model, the entropy floor is related to the cooling time (as
in Figure~\ref{fig-Kcevol}).  This model predicts that $K_\circ$
depends on mass, increasing by a factor of $\sim 5$ from $\sim
0.8K_{100}$ to $\sim 4K_{100}$ over the temperature range of interest.  The resulting
M--L relation is shown as the dashed line in Figure~\ref{fig-ML}; the
slope here is steeper than for any of the fixed
entropy--floor models shown in Figure~\ref{fig-ML2}.  Note, however,
that 
while $K_{\circ}$ may vary by a
factor of $\sim 5$ over this mass range, a similar or larger range of
$K_\circ$ is required to explain the distribution of luminosities at
fixed mass.  Although the cooling model can easily accommodate scatter
toward more luminous clusters (since $K_{\rm cool}$ is only the maximum
entropy of gas that can cool), there is no simple mechanism to account
for the scatter of clusters toward lower luminosities.
Therefore the simple interpretation 
that the entropy threshold is due solely to the cooling of low entropy gas
is not likely to be correct, a conclusion
also reached by \citet{McCarthy-cooling} and \citet{Borgani+05} using more sophisticated
models that account for the 
presence of gas that cools below the $K_{cool}$ threshold.

\subsection{The M-T relation}
\begin{figure}
\leavevmode \epsfysize=8cm \epsfbox{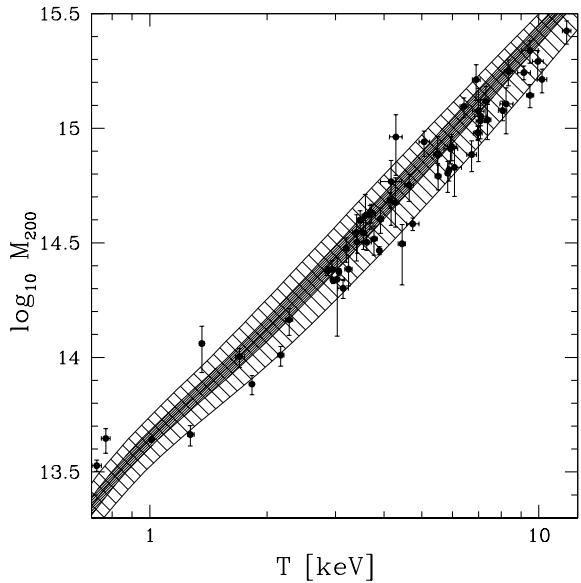}
\caption{The relation between X-ray temperature
  and $M_{200}$. 
The {\it points} 
are local ($z<0.2$) data, with masses from the HIFLUGCS sample
  \citep{HIFLUGCS} and temperatures (uncorrected for cooling flows)
  from \citet{Hornerthesis}. 1$-\sigma$ error bars on the masses are shown; only
  galaxies with relative errors of $<50$ per cent are plotted.  
The 
{\it 
shaded region} represents the
prediction of the shifted-entropy model with $K_\circ=2K_{100}$ and a
realistic 1$\sigma$ (heavy shading) or 3$\sigma$ (lighter shading) scatter in the halo concentration parameter.
\label{fig-MT}}
\end{figure}
In Figure~\ref{fig-MT} we compare the mass-temperature relation
predicted by our models with the \citet{HIFLUGCS} catalogue, again excluding
those clusters with mass
uncertainties greater than 50 per cent, and using temperatures from
\citet{Hornerthesis}.  
The shifted--entropy model provides a good description of the data over
two orders of magnitude in mass, although it is not statistically the
best fit.  

Unlike the M--L relation, the observations here show remarkably little scatter, 
and this scatter is consistent with the published measurement
uncertainties. 
Recently, \citet{GPS4} used strong gravitational lensing to
measure the masses of ten X--ray luminous clusters and found that there is intrinsic scatter
in the M--T relation, due mostly to merging, non-equilibrium systems.
Although this scatter is not apparent in Figure~\ref{fig-MT}, it is
possible that it is
underrepresented here, as $M_{200}$ is obtained
from the X-ray data by
assuming isothermality and imposing a functional form for the surface
brightness profile \citep{Rasia+04}. This parametrization could have the effect of homogenizing 
clusters with a range of cooling core sizes, geometries, and amount of
substructure.  The same would also be true for
the masses in the M--L relation, where we do see significant scatter
(Figure~\ref{fig-ML}), and this could indicate
that the intrinsic scatter in the luminosities is 
larger than the intrinsic scatter in the temperatures.  On the other
hand, the fact that X--ray derived masses are directly proportional to
the temperature introduces a correlation that could reduce the scatter
in the M--T relation alone.  It would be
useful to have a larger sample of clusters with accurate lensing masses
and X--ray observations to improve our understanding of the scatter in
these relations.

\begin{figure}
\leavevmode \epsfysize=8cm \epsfbox{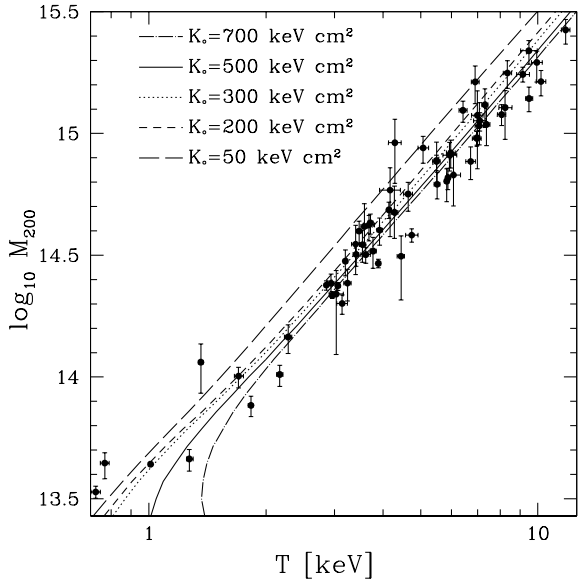} 
\caption{As
  Figure~\ref{fig-ML2}, but for the mass--temperature relation.
\label{fig-MT2}}
\end{figure}
The model predictions in Figures~\ref{fig-MT} and \ref{fig-MT2} show that
the predicted temperature is relatively insensitive to both the halo
structure (i.e. concentration) and the entropy floor, for
$K_\circ>0.5K_{100}$.  Recall that the distribution of halo concentrations
partly arises from substructure, triaxiality and departures from 
equilibrium in simulated clusters and therefore our predicted scatter
approximately includes these effects.  However, in our model these
concentrations are applied to spherical, smooth halos and
the scatter in predicted temperatures is therefore not as large as in numerical
and analytic models that do not make these assumptions \citep{RTK,OBB}.
The insensitivity of our predicted temperatures to the value of the entropy
floor is due to the fact that the higher central
temperature associated with larger $K_\circ$ is offset by the
flattening of the central density profile, which means the
luminosity--weighted temperature is dominated by the temperature at
larger radius.  Thus, even a model with $K_\circ=0.5K_{100}$ predicts
temperatures that are just within the scatter of the observations.  For
the same reason, the $K_{\rm cool}$ model prediction does not differ
significantly from those shown here, so we have omitted it from the
figure for the sake of clarity.

In summary, we have found that a variation of nearly an order of magnitude in $K_\circ$ is required
to explain the large scatter in the $M-L$ relation.  Variations in dark matter halo shape
(concentration) alone are insufficient.  Encouragingly, the wide range
of entropy levels required does not conflict with the small observed scatter in
the $M-T$ relation, because of the temperature's insensitivity to the
core properties.  
Although both theory \citep[e.g.][]{RTK,OBB} and
observations \citep{GPS4} suggest that there may be additional scatter in the M-T relation
that is not apparent under the simplifying assumptions of both the
models and data presented here,
it is still less than the scatter in the M--L relation, which is
dominated by variations in the entropy distribution of the gas.

\section{Scatter and the Luminosity and Temperature Functions}\label{sec-Tfunc}\label{sec-Lfunc}
If the mass spectrum of dark matter halos is known precisely, then the
observed shape and normalization of the temperature and luminosity
functions provides an independent test of the theoretical models, that
does not depend on an observational determination of cluster mass.
Scatter in the mean relations plays an important role here, and can influence the shape of these functions.

We construct the theoretical luminosity and temperature functions 
using the
dark matter mass function of \citet{EvrardHV}, based on the fitting
formalism of \citet{Jenkins},
which provides a universal description of the mass function to within
about 10 per cent.  The advantage of the \citet{EvrardHV} mass function is that
it is expressed in terms of $M_{200}$, the same mass that we use to
compare with the observed mass-temperature and mass-luminosity
relations\footnote{We note that $M_{500}$ is better determined
observationally, and it would be useful to have a mass function from
numerical simulations filtered on this scale.}.
The form
of the temperature and luminosity functions are then completely
determined by the correlation between virial mass and the X-ray
observable (BBLP\footnote{However,
both the observed and model luminosity functions shown in Figure~9 of
BBLP are incorrect due to errors in the bolometric correction and
cosmology conversion.}).  

\subsection{The temperature function}
\begin{figure}
\leavevmode \epsfysize=9cm \epsfbox{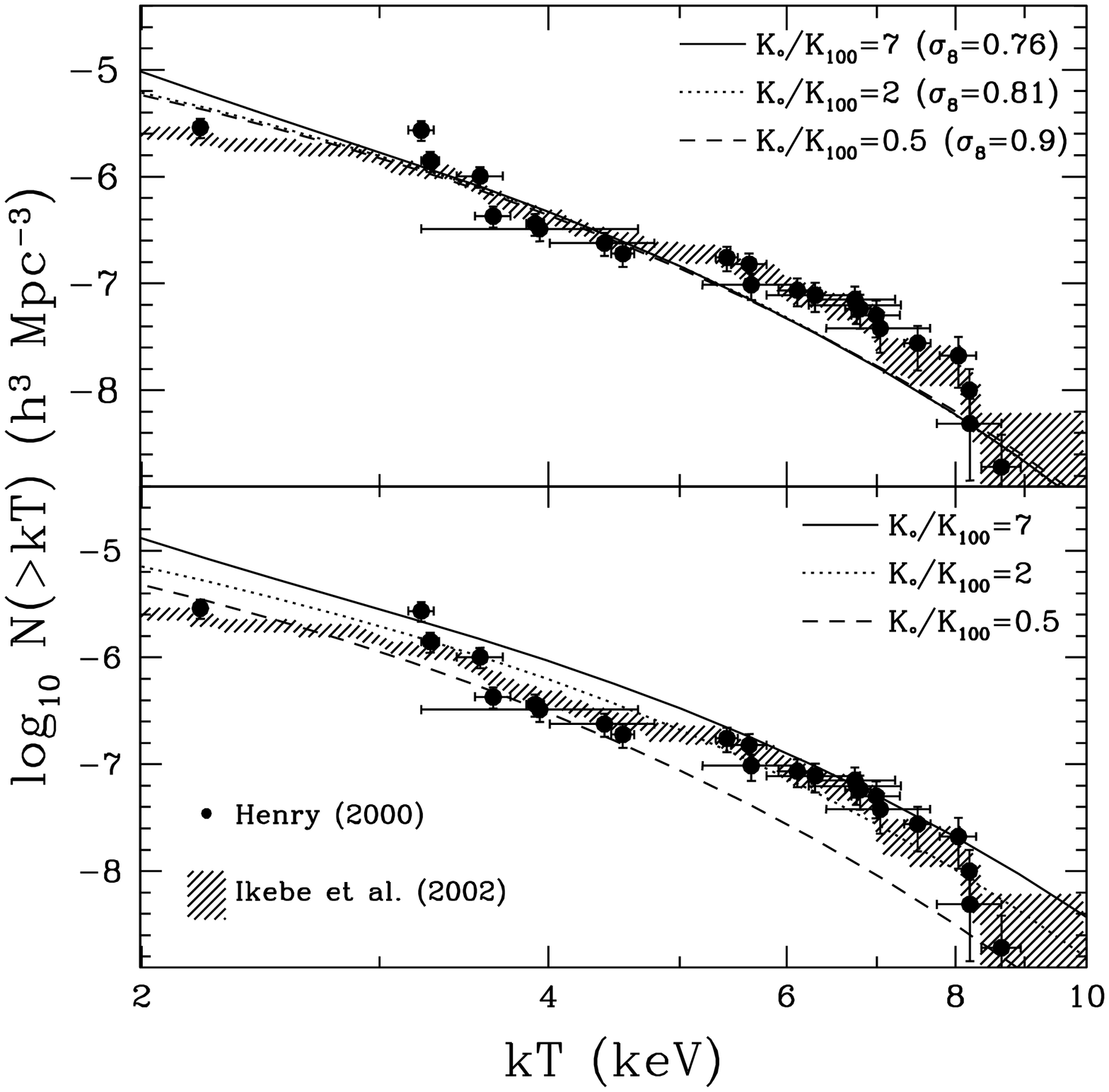}
\caption{The shaded region is the observed temperature function at $z=0$ from \citet{Ikebe}, but with
  temperatures, uncorrected for cooling flows, taken from \citet{Hornerthesis}.  The data points are from \citet{Henry00}.  In the {\it
    bottom panel} we show the shifted-entropy model with three different
  levels of heating,
  assuming $\sigma_8=0.85$ as determined from WMAP \citep{WMAP_short,Tegmark03}.   Models are convolved with a 10 per cent uncertainty in
  temperature which flattens the temperature function.  In the {\it top
    panel}, we show the same models, normalized at $4$--$5$ keV with different values of
  $\sigma_8$, as indicated.
\label{fig-Tfunc}}
\end{figure}
The observed $z=0.15$ temperature functions from \citet{Ikebe} and
\citet{Henry00} are shown in Fig.~\ref{fig-Tfunc}.  For consistency, the temperatures
are taken from \citet{Hornerthesis}, and therefore uncorrected for any
cooling flow component, though this makes little difference in practise.  
To compare with these data, we show
predicted temperature functions from the shifted-entropy models with a
range of normalizations that approximately accounts for the scatter in
the M--L relation.  We
have smoothed the models with 10 per cent Gaussian random noise
on the temperatures, to mimic the scatter in the mean relation (which
is consistent with being due to observational uncertainties). 
First, in the bottom panel, we show three models with different
$K_\circ$ but the
same normalization $\sigma_8=0.85$, as measured from the WMAP data
\citep[and adjusted for the slightly non-concordance values of
cosmological parameters that we have adopted][]{WMAP_short,Tegmark03}.  
At the hot end of the temperature function, the data is best matched by
the models with the highest entropy floors, while at the opposite
extreme the low-entropy models fare better.  However, we caution that 
at low temperatures ($\lesssim 1$ keV) there may be completeness issues that could artificially 
flatten the temperature function and, therefore, yield values of $K_\circ$ 
that are systematically lower than that of the average system \citep[e.g.][]{OP04}.

In the top
panel of Fig.~\ref{fig-Tfunc}, we again show the predicted temperature
functions for a range of $K_\circ$ values, but with $\sigma_8$ 
adjusted to give
the same number density of clusters at $T=4$--$5$ keV, where observational data
from different studies are in best agreement \citep{Ikebe}.  The
best-fit value of $\sigma_8$ is $\sim 3.5$ per 
cent higher if the normalization to the data is made over the range
$kT=6$--$7$ keV.  The range of $K_\circ$ (which are all reasonably
consistent with the mass--temperature relation) corresponds to a range
of best-fit values of $\sigma_8$ ranging from 
$0.76$ to $0.9$.  This is comparable to the observational uncertainty
on this parameter \citep{WMAP_short},
and therefore we cannot use the temperature function alone to provide a
sensitive test of the size of the entropy floor. This is again simply because
temperature is relatively insensitive to the entropy of the central gas.

\subsection{The Luminosity Function}\label{sec-lfunc}
\begin{figure}
\leavevmode \epsfysize=8cm \epsfbox{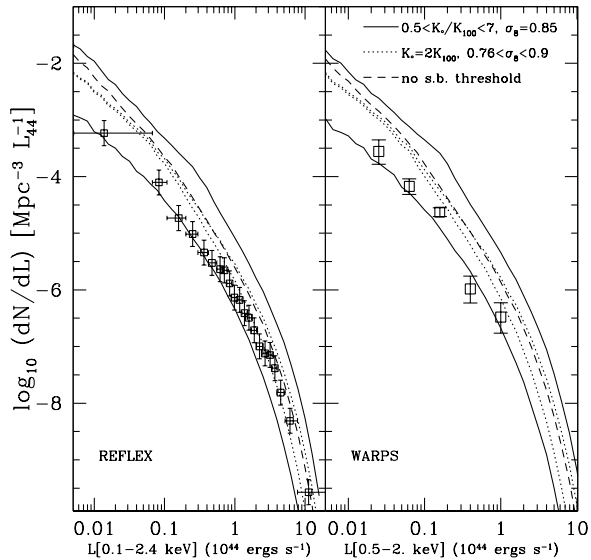}
\caption{
The lines in each panel show different theoretical models for the
luminosity function. The {\it solid lines} show models with entropy
floors of $K_\circ=0.5K_{100}$ (upper line) and $K_\circ=7K_{100}$
(lower line).  This approximately brackets the range of entropies
required to explain the scatter in the M-L relation
(Figure~\ref{fig-ML}). The {\it dotted lines} show the effect of
varying $\sigma_8$ as indicated, keeping $K_\circ=2K_{100}$ fixed;
lower values of $\sigma_8$ reduce the number of luminous clusters by a
small amount but have no effect on the faint end of the luminosity
function.  Finally, the {\it dashed line} shows the $K_\circ=2K_{100}$
model but 
  omitting the surface brightness threshold, which increases the
  prediction at low luminosities.    
{\it Left panel:} The {\it open squares} with error bars are
  the observed local luminosity function from the REFLEX survey
  \citep{B+01_short}.  {\it Right panel:} Similar, but
  where the data are from the WARPS
  \citep[][Jones et al., in prep.]{WARPS_LF}, in a different
  energy band. 
\label{fig-Lfunc}}
\end{figure}
To compute the luminosity function, the intrinsic scatter in the M-L
relation must be taken into account. Because of the steepness of the
mass function, even a small distribution of halo masses corresponding
to a given luminosity can have an important effect on the number
density of clusters at that luminosity.  Unfortunately, the value of
$K_\circ$ in our model does not have a 
unique physical motivation, and thus we do not have a prediction for
the scatter as a function of mass.  However, we can see from the data
in Figure~\ref{fig-ML} that the observations are approximately covered by
models with a range of entropy floors $0.5<K_\circ/K_{100}<7$, so in
Figure~\ref{fig-Lfunc} we
show the prediction of the luminosity function for these two
extremes.  The true luminosity function should lie between these
limits, with a shape that depends on the distribution of entropy levels at each luminosity.

The observed luminosity functions at $z=0.15$ from the 
WARPS \citep[][Jones et al., in prep.]{WARPS_LF} and REFLEX
\citep[][]{B+01_short} surveys are shown in
Figure~\ref{fig-Lfunc}.  
The solid lines show our model, for $K_\circ=0.5K_{100}$ (upper line)
and $K_\circ=7K_{100}$ (lower line).  This range brackets the
observational data, although the data do lie nearer the model with high
entropy.  This is especially true for the low-luminosity clusters, with
$L\lesssim10^{44}$ergs~s$^{-1}$.  This may indicate that lower--mass
clusters have higher central entropies, on average; this may also be
evident from Figure~\ref{fig-ML}, although there are few clusters with
accurate mass measurements at these low luminosities.

We also show, as the dotted lines, the effect of varying 
$\sigma_8$ between 0.76 and 0.9 (assuming $K_\circ=2K_{100}$).  This
has only a small effect on the bright end of the luminosity function, and no effect on the faint end.
The dashed line shows the effect of
removing the limiting surface brightness threshold of $1\times 10^{-15}$
ergs~s$^{-1}$~cm$^{-2}$~arcmin$^{-2}$ \citep{WARPSI} used in the other
models; these surface brightness corrections are relevant
only for the least luminous clusters in the sample.

Thus we have shown that consistency between the X--ray scaling
relations (M--L and M--T) and the luminosity function can be achieved
in these models; however, a better understanding of the entropy--floor
distribution as a function of mass is required to make a firm
prediction of the luminosity function shape.   

\section{Evolution}\label{sec-evol}
We now turn to the redshift evolution of the X-ray scaling
relations and the temperature and luminosity functions.  These
predictions can provide another interesting test of the difference
between the 
fixed--$K_\circ$, preheating models and models where $K_\circ=K_{\rm
  cool}$. 

\begin{figure}
\leavevmode \epsfysize=8cm \epsfbox{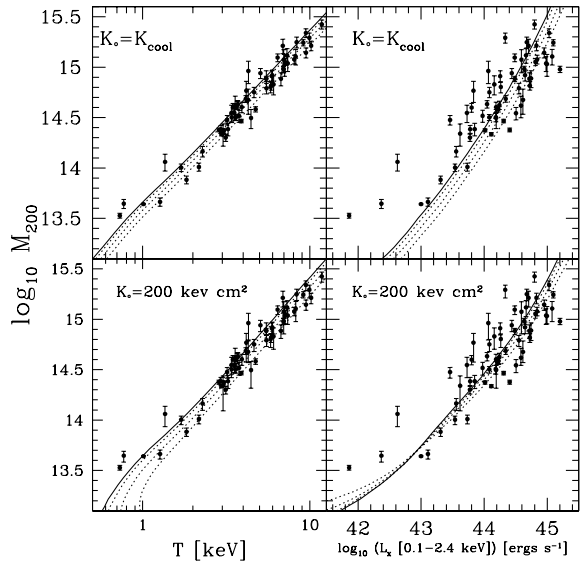}
\caption{The observed M--L and M--T relations shown in
  Figs.~\ref{fig-ML} and \ref{fig-MT} are reproduced as the {\it solid
    circles}.  In the {\it bottom panels} we show the default
  model predictions, where the entropy floor is independent of redshift, at $z=0$ ({\it solid line}) and $z=0.15, 0.4, 0.7$
  ({\it dotted lines}).  The curves in the {\it top panels} are the
  models where the entropy floor $K_{\rm cool}$ is related
  to the cooling time of the gas and thus evolves with redshift.
\label{fig-MTLevol}}
\end{figure}

In Fig.~\ref{fig-MTLevol} we show the predicted evolution in the mass-temperature and mass-luminosity
relations for the two models.  
The data are the same $z\sim 0.15$
data shown in Figures ~\ref{fig-ML} and \ref{fig-MT}, and  the model
predictions are shown at $z=0,0.15,0.4$ and $0.7$.  
For massive
clusters, $T\gtrsim 4$ keV, the predicted evolution in both
the M--T and M--L correlations is
mostly in the normalization, with little change in the slope.

The amount of evolution in the M--L relation is small
relative to the observed scatter at $z=0$.  
On the other hand predicted
evolution in the M--T relation is
more noticeable; clusters at a given temperature are predicted to be
40--60 per cent less massive at $z\sim 0.7$.  The sense and magnitude 
of the evolution are comparable to recent {\it XMM--Newton} and {\it
  Chandra} data \citep{KV05,Maughan+05}.
Both the
$K_\circ=2 K_{100}$ and the $K_\circ=K_{\rm cool}$ models models predict a
similar amount of evolution for the $M-T$ relation, so this is
not a useful way to discriminate between them.  

\begin{figure}
\leavevmode \epsfysize=8cm \epsfbox{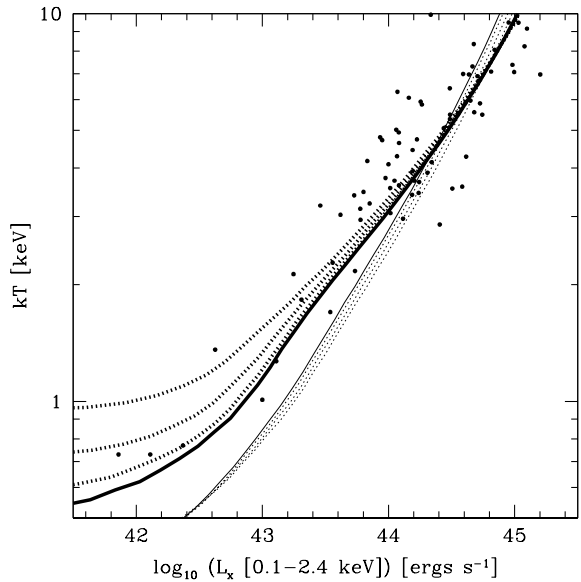}
\caption{The observed temperature--luminosity relation at $z=0$ is shown as the {\it solid
    circles}.  The {\it thin} and {\it thick} lines represent the
  $K_{\rm cool}$
  and $K_\circ=2K_{100}$ models, respectively.  Predictions are shown for $z=0$ ({\it solid line}) and $z=0.15, 0.4, 0.7$
  ({\it dotted lines}).  
\label{fig-LTevol}}
\end{figure}

Interestingly, although the evolution in the M--T and M--L relations
appear similar for both models, the predicted evolution of the L--T scaling law is in
opposite directions, as shown in Figure~\ref{fig-LTevol}.  At high
temperatures, the amount of evolution in the predicted relation is very small.
For the fixed--floor model, the evolution is negligible, while
the $K_{\rm cool}$ model predicts that high redshift
clusters will be about 30 per cent more
luminous at fixed temperature, due to the fact that the entropy floor
is lower at higher redshift.  Observations of distant clusters
seem to indicate a much stronger evolution, with clusters at $z\sim
0.7$ being up to 2.5 times brighter than local clusters of the same
temperature \citep{Vik+02,Lumb+04,KV05,Maughan+05}.  In principle, this
observation has the potential to rule out both models shown here.
However, we note that the amount of evolution observed is sensitive to
the details of the analysis  \citep[e.g.][]{Ettori04_2,Ettori04}.   Since
any evolution in the mean scaling relation is much less than the factor
$\sim 4$ scatter in luminosity at fixed temperature for local clusters,
it is probably premature to claim these observations rule out either
model until the scatter and the selection biases that result from it
(i.e. Malmquist--like bias) are robustly integrated into the model predictions.
The most leverage will come from low temperature systems at high
redshift; the $K_\circ=2K_{100}$ model predicts very strong (negative) evolution in the
luminosities of these groups, as the fixed entropy floor becomes very
large relative to the characteristic entropy.

\begin{figure}
\leavevmode \epsfysize=8cm \epsfbox{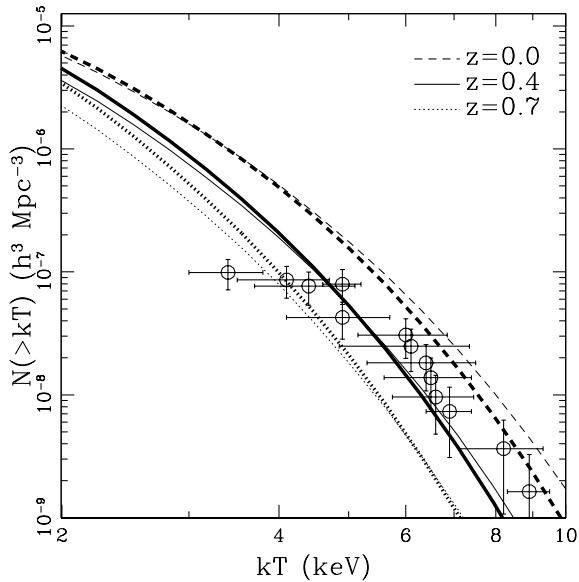}
\caption{The predicted temperature functions at $z=0, 0.4$ and $z=0.7$, for
  the fixed--entropy models ({\it thick lines}) and the $K_{\rm cool}$ models ({\it thin lines}).  The data are  at $z\sim 0.4$, from \citet{Henry00}.  
\label{fig-Tfunc_hiz}}
\end{figure}

The evolution of the temperature and luminosity
functions requires a knowledge of the mass function at $z>0$.  Since
this has not yet been precisely measured from simulations, we take the
local mass function from \citet{EvrardHV} and evolve it to higher
redshift using Press-Schechter theory.
In Figure~\ref{fig-Tfunc_hiz} we show the temperature function for 
both models (normalized at $kT\sim$4--5 keV, with $\sigma_8\sim0.8$), at $z=0, 0.4$ and $z=0.7$,
compared with $z=0.4$ data from \citet{Henry00}\footnote{An updated version of the observed high-redshift temperature function, based on more data, is presented in \citet{Henry04}.  The results are consistent with those shown here.}.  Over the range of observed temperatures, $kT>3$
keV, the evolution in the two models is similar, and in good agreement
with the observations.  The models begin to diverge at lower
temperatures, where the fixed--entropy model predicts a little less evolution.

\begin{figure}
\leavevmode \epsfysize=8cm \epsfbox{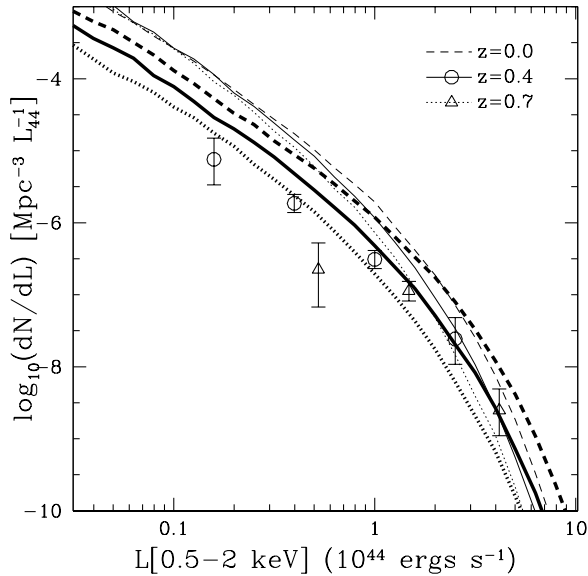}
\caption{The luminosity functions at $z=0, 0.4$ and $z=0.7$,  for
  the fixed--entropy ($2K_{100}$) models ({\it thick lines}) and
  the $K_{\rm cool}$ models ({\it thin lines}). Data at $z=0.4$ and $0.7$ are shown, from the WARPS
  survey, updated with {\it XMM-Newton} and {\it Chandra} luminosities
  for clusters at $z>0.6$.
\label{fig-Lfunc_hiz}}
\end{figure}

The predicted evolution of the luminosity function is shown in
Figure~\ref{fig-Lfunc_hiz}.  Unfortunately, our model does not predict
the scatter in the M--L relation nor its evolution, so for illustration
we have just shown the $K_\circ=2K_{100}$ model, which provides a
reasonable match to the bright end of the local luminosity function.
The models are again compared with
data at similar redshifts, from the WARPS survey \citep[][Jones et al.,
in prep.]{WARPS_LF}.  The WARPS data at z$>$0.6 have been updated with
accurate  
luminosities measured from XMM-Newton and Chandra observations, and are
corrected to the rest-frame energy range 
0.5--2.0 keV. 
In contrast with the
temperature function, the observed luminosity function evolution is modest, with a factor $\sim3$
decrease in the number of massive clusters between $z=0$ and
$z=0.7$. 

The $K_{\rm cool}$ model predicts very little evolution in the
luminosity function,
and thus overpredicts the number of high redshift clusters.
The $K_\circ=2K_{100}$ model appears to be in much better
agreement with the data, especially for the brightest clusters.  As with the local luminosity function, the
disagreement at the faint end may indicate that lower luminosity
clusters have larger central entropies.  However, as we have already seen (Figure~\ref{fig-Lfunc}), the
shape of the luminosity function is very sensitive to the amount of
scatter in the M--L relation, and we have no theoretical or empirical
knowledge about how this scatter evolves.
We also note that the high--luminosity end of the luminosity function
is still poorly determined, with a variation in observed number abundance at
fixed luminosity measured from different surveys at $z>0.3$ 
being about a factor $\sim 2$ \citep{Mullis04}.  

\section{Discussion and conclusions}\label{sec-discuss}\label{sec-conc}
In this paper, we have revisited the observed X-ray scaling
relationships between mass and temperature (M--T) and mass and
luminosity (M--L), simultaneously with the temperature and luminosity
functions. We have attempted to compare the observable quantities as
directly as possible (i.e. without bolometric or cooling--flow
corrections) with a simple suite of models in which a fiducial gas
entropy distribution is shifted by a value $0.5<K_\circ/K_{100}<7$.
In particular, we focus on the scatter in the observed scaling
relations, and how this compares with the scatter expected due to a) a
range of halo structures (concentrations); b) the time available for
cooling--only processes or c) heating/cooling
efficiency.  Since cluster temperatures are relatively insensitive to 
variations in the entropy distribution (and hence the scatter in the M--T relation is small), we
gain the most by focusing on the M--L relation and the luminosity
function. 
Our main findings are as follows:
\begin{enumerate}
\item The variations in dark matter halo concentration expected from
  simulations are not
  large enough to account for the scatter in the observed M--L
  relation of clusters and groups.
\item Simple models of the intracluster medium in which the core
  entropy is modified to have a minimum value require the value of
  this floor to be  between about 0.5$K_{100}$ and $7K_{100}$ to match the slope, normalization
  and scatter in the observed M--T and M--L scaling relations.  The
  constraint comes mostly from the M--L relation, as the temperature is
  insensitive to the value of $K_\circ$.
\item The shape of the luminosity function is sensitive to the scatter
  in the M--L relation.  
The observations lie between the models with
$K_\circ=0.5K_{100}$ and $K_\circ=7K_{100}$, but closer to the
higher--entropy model.  
  The scatter  in entropy levels as a function of halo mass
must be accounted for if the parameters
  $\sigma_8$ or $K_\circ$ are to be accurately deduced from the luminosity
  function alone.
\item The model temperatures are in good agreement with the observed
  temperature function, assuming the mass function of
  \citet{EvrardHV}.  However, the insensitivity of temperature to
  $K_\circ$, and the uncertainty on the normalization parameter
  $\sigma_8$, means this does not put strong constraints on the value
  (or range of values) of $K_\circ$.
\item The amount of gas that can cool in a Hubble time sets a maximum
  value on the minimum entropy of intracluster gas, $K_\circ$.  Lower
  entropy floors can be achieved by allowing some gas to cool below the
  threshold, or reducing the time available for cooling.  However,
  although this model provides a reasonable match to the median
  mass--luminosity relation, it cannot account for the many clusters
  with luminosities below this relation.
\item We also present predictions for the evolution of the scaling relations and
  temperature/luminosity functions, out to $z=0.7$.  A comparison with observations suggests that the 
model with $K_\circ=2K_{100}$ is valid out to $z\sim 0.7$, but
this depends on how the scatter in the M--L relation evolves, which is
currently unknown. On
  group scales, where observations unfortunately will be most difficult, these predictions 
  clearly distinguish between an entropy floor $K_\circ$ that is
  independent of mass and redshift, and one that is tied to the cooling
  time of a halo.  
\end{enumerate}

Thus, the scatter in the observations can be understood if the minimum entropy
of the gas in clusters has a median value of $K_\circ=200$ keV~cm$^2$ but varies by a
factor $\sim 3$ between halos of similar mass.  Although we lack a
quantitative theoretical prediction for the origin of this scatter, it
seems reasonable to associate it with a similar range in the
efficiency of heating and/or cooling.

We acknowledge that  the models presented here provide an
incomplete description of the intracluster medium.  Apart from the
obvious point that the pure heating models neglect the cooling
processes that must take place to form the galaxies and AGN responsible
for the heating in the first place, these models also lead to isentropic core gas
distributions that are in conflict with at least some high resolution
observations of clusters \citep[e.g.,][]{PSF,PA04}.  Therefore, cooling
must be incorporated for a full description of the data.  Although we
have made a very crude step in this direction by tying the entropy
floor level to the cooling time available, this model is greatly
oversimplified, as it does not allow the gas 
to flow to the centre as it cools as in more realistic models (e.g. VBBB).
Furthermore, for low mass clusters
and groups this model predicts too much condensed gas, as most of the
intracluster medium in a self--similar model has an entropy below the
cooling threshold. Therefore, some combination of heating and cooling
is expected to be required.  

Such a model has recently been developed
by \citet{McCarthy-cooling}; in this model, preheated gas is allowed to
cool in a realistic way, following the hydrodynamic evolution of the
gas as it flows to the centre.  With an appropriate range of preheating
levels and cooling times, this model can reproduce the slope,
normalization and the scatter in the M--L relation.  Our results show
that including scatter in the halo potentials (not considered by McCarthy et al.)
would only have a secondary impact on their constraints.   We have also
explored the effect of scatter on the temperature and luminosity
functions.  Since
the M--T relation is relatively insensitive to the gas
distribution (as long as the very lowest entropy gas is heated or
removed), we can expect that the predictions for the temperature
function using the model of McCarthy et al. would be very similar to
the models presented here.  However, the luminosity 
function, and the evolution in the scaling relations, will be sensitive
to the mass and redshift dependence of the distributions of preheating
levels and cooling times.  The next step, therefore, is to physically
link these distributions through a feedback model, and to compare the
resulting luminosity function with the observations.

Finally, we note that while analyses of the M--L and M--T relations and 
the luminosity and temperature functions can teach us much, they are 
only probing the integrated properties of the intracluster medium.  Some of the most 
powerful constraints on the role of non-gravitational physics in mediating the 
properties of the ICM are likely to come from detailed comparisons of 
theoretical models to the actual spatially-resolved profiles (e.g., entropy 
and temperature) of clusters.  The difficulty is in obtaining such 
profiles for large, representative samples of clusters and (especially) 
groups.  However, as more and more groups and clusters observed with {\it 
Chandra} and {\it XMM-Newton} become publicly available, progress is beginning 
to be made on this front \citep[e.g.,][]{PA04, Piff05,C+05}.  Such comparisons will 
make excellent complementary probes to studies such as the present one that 
make use of more easily obtained and robust integrated cluster properties.

\section*{Acknowledgements}
We thank D. Horner for providing his cluster catalogues in electronic
form, and we acknowledge useful discussions with Scott Kay and Paul Bode.
MLB is grateful for the hospitality at University of Victoria, where
this work was initiated, and acknowledges financial support from a
PPARC fellowship PPA/P/S/2001/00298 and an NSERC Discovery grant.
Research support for AB comes from the Natural Sciences and Engineering 
Research Council (Canada)
through the Discovery grant program.   AB would also like to acknowledge 
support from the Leverhulme
Trust (UK) in the form of the Leverhulme Visiting Professorship at the 
Universities of Oxford and Durham.  IGM acknowledges
support from an NSERC postgraduate scholarship.
\bibliography{ms}
\end{document}